\documentclass[runningheads]{llncs}
\usepackage{graphicx}
\usepackage{amsmath}
\usepackage{subfigure}
\usepackage{dsfont}

\usepackage{siunitx}

\begin{document}
\title{{Translating Simulation Images to X-ray Images via Multi-Scale Semantic Matching}}

\author{Jingxuan Kang\inst{1} \and
Tudor Jianu\inst{1} \and
Baoru Huang\inst{2}  \and
Binod Bhattarai\inst{3} \and
Ngan Le\inst{4} \and
Frans Coenen \inst{1}  \and
Anh Nguyen \inst{1}}
\institute{Deparment of Computer Science, University of Liverpool, UK \\
\email{sgjkang3@liverpool.ac.uk} \and
The Hamlyn Centre for Robotic Surgery, Imperial College London, UK  \and
University of Aberdeen, UK \and
University of Arkansas, USA
\\}
\maketitle   

\begin{abstract}
Endovascular intervention training is increasingly being conducted in virtual simulators. However, transferring the experience from endovascular simulators to the real world remains an open problem. The key challenge is the virtual environments are usually not realistically simulated, especially the simulation images. In this paper, we propose a new method to translate simulation images from an endovascular simulator to X-ray images. Previous image-to-image translation methods often focus on visual effects and neglect structure information, which is critical for medical images. To address this gap, we propose a new method that utilizes multi-scale semantic matching. We apply self-domain semantic matching to ensure that the input image and the generated image have the same positional semantic relationships. We further apply cross-domain matching to eliminate the effects of different styles. The intensive experiment shows that our method generates realistic X-ray images and outperforms other state-of-the-art approaches by a large margin. We also collect a new large-scale dataset to serve as the new benchmark for this task. Our source code and dataset will be made publicly available.


\keywords{Sim2Xray \and GAN \and Interventional Simulation Systems}



\end{abstract}

\section{INTRODUCTION} \label{Sec:Intro}
Image-to-image translation involves converting an image into a different modality or style~\cite{fu2019geometry,zheng2021spatially,zhu2017unpaired}. In medical imaging, this task is related to the translation between various medical image modalities, such as MRI to X-ray~\cite{stimpel2019projection}, MRI to CT~\cite{nie2017medical}, or between MRI modalities \cite{bui2020flow}. Medical image translation is challenging due to the need for preserving semantic and structural information, as well as the details during the translation process. In practice, medical images often share similarities, with only minor differences. Effective translation methods can significantly aid medical training~\cite{armanious2020medgan,yan2022swin}, surgical planning~\cite{sharan2021mutually}, or sim-to-real learning~\cite{Paavilainen_2021}. However, challenges such as data privacy and incompleteness hinder medical image transfer, while deep learning algorithms require extensive data, compounding these issues~\cite{https://doi.org/10.48550/arxiv.1908.10454}.

The recent development of surgical simulators~\cite{https://doi.org/10.48550/arxiv.2208.01455,sutherland2006surgical} facilitates the acquisition of medical skills of aspiring surgeons. Compared to real-world setup, training learning algorithms in simulation is inexpensive and expeditious~\cite{kunkler2006role,tran2022light}. However, most of the current medical simulators consider gray-scale as X-ray images. This assumption causes a challenging problem when we apply the learned knowledge from the medial simulators to the operating theater~\cite{dagnino2022vivo}. To bridge the gap between simulation images from medical simulators and real medical images, several works have proposed GAN-based methods for medical image translation through adversarial training~\cite{bermudez2018learning,frid2018gan,nie2017medical}. However, these methods usually have the collapsed pattern problem or fail to yield valuable results~\cite{kodali2017convergence}. 

\begin{figure}[t]
\centering
\subfigure[]{\includegraphics[width=0.28\linewidth, height=0.28\linewidth]{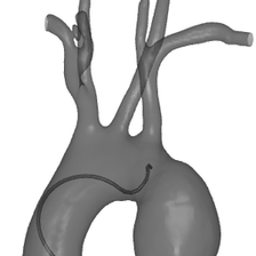}}
\hspace{1px}
\subfigure[]{\includegraphics[width=0.28\linewidth, height=0.28\linewidth]{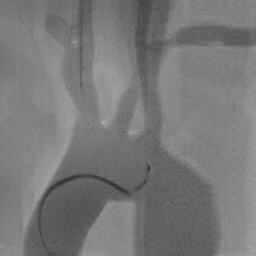}}
\hspace{1px}
\subfigure[]{\includegraphics[width=0.28\linewidth, height=0.28\linewidth]{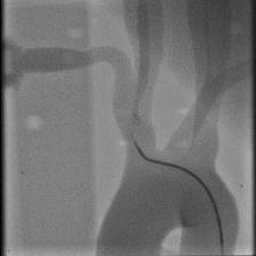}}
\caption{We present a new method to translate the images from an endovascular simulator to X-ray images. Our method preserves the \textit{structure information of the input simulation image} and learns the \textit{``X-ray style" from the real X-ray image}. (a) input image (b) our generated image, (c) an example of a real X-ray image.}
\label{fig:intro}
\end{figure}

In this paper, we propose a simple, yet effective method translate \textit{unpaired simulation images} from an endovascular simulator~\cite{https://doi.org/10.48550/arxiv.2208.01455} into X-ray images. Unlike previous works that focus on X-ray images with clear and static human body parts (e.g., X-ray images of the hand)~\cite{Haiderbhai2020pix2xrayCR}, our input are endovascular simulation images which contain dynamic motion of the catheter~\cite{nguyen2020end,kundrat2021mr}. Therefore, we need to learn both the ``style" information of the real X-ray image, while maintaining the \textit{structure of the input} (e.g., the position of the catheter). To this end, we introduce a multi-scale domain matching method to learn both the style and preserve the structure information during the translation. As shown in Fig.\ref{fig:intro}, our model archives realistic results that are almost indistinguishable between the translated image and the real X-ray image.  
Furthermore, our model's simple architecture enables rapid training and inference, making it well-suitable for real-time endovascular simulators. Additionally, we introduce a new and challenging dataset of unpaired images, consisting of $1,607$ real X-ray images and $2,000$ simulation images. This dataset is essential for developing and evaluating robust image translation models, particularly in the under-studied task of translating simulation images to real X-ray images.

\section{Related Work} \label{Sec:rw}

\subsection{Image to Image Translation}
Numerous works have focused on conditional GAN~\cite{arjovsky2017wasserstein,frid2018gan,zhang2017stackgan} for image-to-image translation since the introduction of GAN~\cite{goodfellow2020generative}. CycleGAN~\cite{zhu2017unpaired} proposed a cycle consistency loss to constrain the model, allowing it to translate an image back to the original domain after being translated to the target domain.  Another well-known work was GcGAN \cite{fu2019geometry}. The limitation of CycleGan and GcGAN was the lack of clear constraints on the process of converting, which might produce multiple solutions thus, not meeting the requirements of medical image translation. DistanceGAN~\cite{benaim2017one} solved the model collapse problem, but it did not impose any constraints on semantic information. F-Lesim~\cite{zheng2021spatially} used self-similarity to define the structure of the scene, but it had strong constraints, making it difficult to train and risking the loss of useful semantic information. 

\subsection{Medical Image Translation}
An initial implementation of GAN-based for medical image translation was introduced in~\cite{frid2018gan} for synthesizing different classes of lesion patches of liver CT images~\cite{radford2015unsupervised}. Furthermore, given that CT imaging puts patients at risk of cellular damage and radiation from cancer, CAGAN~\cite{nie2017medical} implemented pixel-by-pixel reconstruction loss and image gradient loss to synthesize CT images from MR images. Nevertheless, it required one-to-one correspondence with the dataset for training. The subsequent Deep MR to CT Synthesis~\cite{wolterink2017deep} used unpaired data and get acceptable results.
MedGAN~\cite{armanious2020medgan} utilized a discriminator network as a trainable feature extractor to penalize differences between the translated medical image and the desired modality. Stylistic transfer loss was used to match the texture and fine structure of the desired target image to the translated image. Based on the theory of loss correction~\cite{patrini2017making},  RegGAN~\cite{kong2021breaking} assumed that aligned data could be treated as noisy labels, and an additional alignment network on the generator could adaptively fit this noisy distribution. Compared to the above works, we directly translate the simulation images into X-ray images without the need for paired data.

Many methods have been proposed to translate RGB images directly to medical images~\cite{amirrajab2022sim2real,Haiderbhai2020pix2xrayCR}. Pix2xray~\cite{Haiderbhai2020pix2xrayCR} utilized CGANs to generate synthetic X-rays. However, obtaining the required dataset for pix2xray is time-consuming as it necessitates RGB images, pose images, and X-ray images. Our proposed method, on the other hand, only requires simulation images that do not need to be paired with real X-ray images. In addition to pix2xray, other approaches such as GDR~\cite{suzuki2022goldilocks} used domain randomization to synthesize realistic images. The authors in~\cite{amirrajab2022sim2real} proposed a method for Cardiac MRI simulation-to-real translation using unsupervised GAN. 


\section{Method} 

Given a collection of simulation images $\mathcal{X}$ and real X-ray images $\mathcal{Y} $, 
our goal is to find a generator $\mathcal{G}$ mapping $\mathcal{X}$ domain to $\mathcal{Y}$ domain, denoted as $ \mathcal{G}: \mathcal{X}\to \mathcal{Y} $. The translated result is $ \hat{\mathbf{y}} =\mathcal{G}(\mathbf{x})$, $\mathbf{x} \in \mathcal{X}$. We aim to convert unpaired simulation images into X-ray images. Due to the demand for keeping details of the input simulation images, we need to maintain the semantic and structured information during the translation process, while changing the style of the input simulation image to the style of the X-ray image.

\begin{figure*}[t]
\centering 
\includegraphics[width=1.0\textwidth]{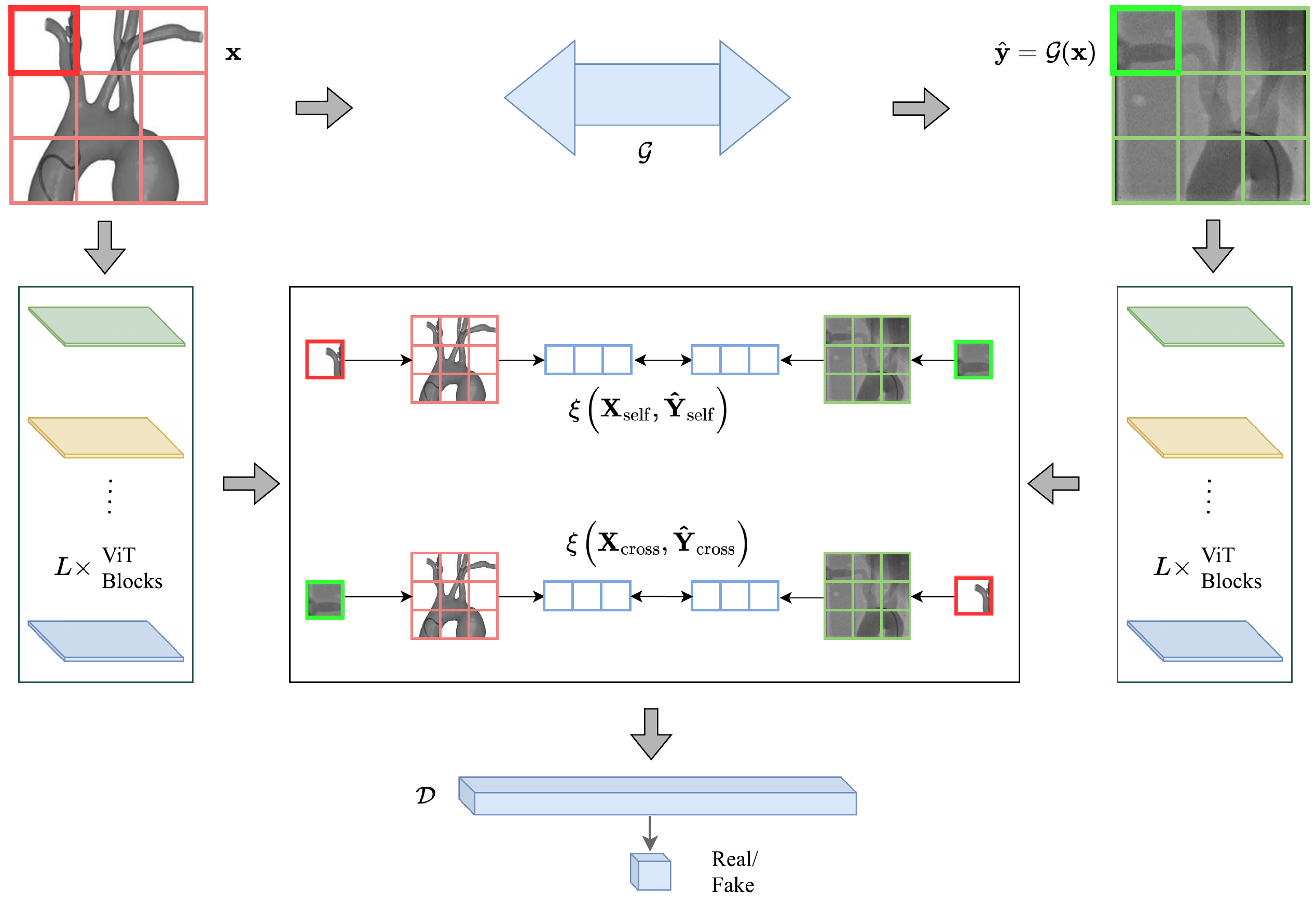} 
\caption{An overview of our framework. We feed the simulation images to the Generator 
to obtain the translated images, and two images enter the same pre-trained ViT network. The features are extracted from the intermediate blocks of the ViT. We match the self-domain and cross-domain respectively to maintain the structure information of the input and learn the style of the X-ray image. Finally, a discriminator is used to classify the fake and real images.}

\label{model} 

\end{figure*}
As shown in Fig.~\ref{model}, we input a simulation image $\mathbf{x}\in \mathcal{X}, \mathbf{x} \in \mathds{R}^{\rm H \times \rm W \times \rm C}$ and use adversarial training to generate an X-ray image $\hat{\mathbf{y}}$. The pre-trained ViT network~\cite{dosovitskiy2020image} extracts high-level structural features by splitting the image into patches that act as tokens. Our multi-scale semantic matching approach maps and learns structural information between the input and output by simultaneously matching a query token to all tokens at other positions of the image. The final semantic matching result is a weighted average of different blocks matched. A single multilayer perception layer (MLP) that takes the features from the ViT is used as the discriminator $\mathcal{D}$ to classify the fake and the real input. 



\subsection{Multi-Scale Semantic Matching}

The adversarial training can learn domain mapping but may generate random permutations of the target domain~\cite{kodali2017convergence}, hence changing the structure of the input image. To address this, we propose Multi-Scale Semantic Matching to maintain semantic structure information relationships and reduce the diverse style impact during adversarial training.

\textbf{Feature Extractor.}
We extract features from simulation images using a pre-trained ViT model~\cite{dosovitskiy2020image} that divides images into patches. Each patch is treated as a query token, with all patches serving as key tokens. We select the output of multiple intermediate blocks and match them separately between domains. 
In practice, we find out that the transformer-based network ViT is particularly well-suited for the simulation images in our problem, which mainly feature catheters and guidewires that span the entire image but occupy a relatively small number of pixels~\cite{han2022survey}. 



\textbf{Multi-Scale Self-Domain Matching.}
We maintain semantic information on multiple intermediate blocks of ViT~\cite{dosovitskiy2020image} to keep the structure information at different scales. To make the simulator image and the generated X-ray image have the same structure relationship, we perform the matching between all tokens in $\mathbf{x}$ and $\hat{\mathbf{y}}$, respectively. We call it \textit{self-domain} matching as the tokens are matched \textit{within the same image}. We use $\mathbf{s
}_i$ to denote the i-th token in $\mathbf{x}$, and $\mathbf{s}_*$ to denote all the tokens in $\mathbf{x}$. Similarly, we denote $\mathbf{t}_i$ as the i-th token in the $\hat{\mathbf{y}}=\mathcal{G}(\mathbf{x})$, and $\mathbf{t}_*$ to denote all the tokens in $\hat{\mathbf{y}}$. Each token can also be matched with itself. We formulate this process as follows:

\begin{equation}
\mathbf{v}_{s_{i}}=\mathbf{s}_{i} \cdot \mathbf{s}_{*}
\end{equation}
\begin{equation}
\mathbf{v}_{t_{i}}=\mathbf{t}_{i} \cdot \mathbf{t}_{*}
\end{equation}
where $ \mathbf{s}_i,\mathbf{t}_i \in \mathds{R}^{d} $, and $\mathbf{s}_*,\mathbf{t}_*\in \mathds{R}^{ n \times d}$. $n$ represents the number of tokens. We match the query token to all tokens to obtain a vector $\mathbf{v}\in \mathds{R}^{n} $. We repeat this process for each query token in the input image to obtain the matrix which contains the semantic relationships between all tokens in the image. We express the matrices as $\mathbf{X}_\mathrm{self}=[\mathbf{v}_{s_{1}},\mathbf{v}_{s_{2}},...,\mathbf{v}_{s_{n}}]$,$\mathbf{\hat{Y}}_\mathrm{self}=[\mathbf{v}_{t_{1}},\mathbf{v}_{t_{2}},...,\mathbf{v}_{t_{n}}]$
. The aim of this process is to achieve semantic alignment between the simulation image and the X-ray image. This is achieved by minimizing the distance between the two $\mathbf{X}_\mathrm{self}$, $\mathbf{\hat{Y}}_\mathrm{self}$  matrices.


\textbf{Multi-Scale Cross-Domain Matching.} The use of self-domain matching does guarantee similar semantic relationships, but the process is inevitably interfered with by stylized information~\cite{kolkin2019style}. The images from the simulator and the generated X-ray style images have completely different style information, which influences the effectiveness of the translation. To avoid the intervention of style information~\cite{chen2021diverse}, we propose multi-scale cross-domain matching to decouple the content and the style. As shown in Fig.~\ref{model}, we match a \textit{token from the simulation image with all tokens from the X-ray image} and vice versa. This match results not only contains the gap between different tokens, but also the gap between different styles. Similarly, we apply this process to all tokens to obtain two semantic representation matrices. In contrast to the self-domain matching, the two matrices have information gaps from different styles. By optimizing the disparity between these two matrices, the effect of style information can be reduced. Specifically, we use the tokens in $\mathbf{x}$ to match with all the tokens in $\hat{\mathbf{y}}=\mathcal{G}(\mathbf{x})$. The matching semantic information contains the gap between different positions and the difference between styles. Then we use the tokens in $\hat{\mathbf{y}}$ to match with all the tokens in $\mathbf{x}$. We formulate this process as follows:
\begin{equation}
 \mathbf{u}_{s_{i}}=\mathbf{s}_{i} \cdot \mathbf{t}_{*}
 \end{equation}
\begin{equation}
\mathbf{u}_{t_{i}}=\mathbf{t}_{i} \cdot \mathbf{s}_{*}
\end{equation}
where $\mathbf{s}_i,\mathbf{t}_i \in \mathds{R}^{d} $, and $\mathbf{s}_*,\mathbf{t}_*\in \mathds{R}^{ n \times d}$. $n$ is the number of tokens.
The two matrices are expressed as $\mathbf{X}_\mathrm{cross}=[\mathbf{u}_{s_{1}},\mathbf{u}_{s_{2}},...,\mathbf{u}_{s_{n}}]$,
$\mathbf{\hat{Y}}_\mathrm{cross}=[\mathbf{u}_{t_{1}},\mathbf{u}_{t_{2}},...,\mathbf{u}_{t_{n}}]$.

\subsection{Training}

We express the multi-scale domain matching objective as follows:

\begin{equation}
 \mathcal{L}_{\mathrm{self}}= \frac{1}{N} \sum_{i=1}^{{N}} \xi \left(\mathbf{X}_{\mathrm{self}},\mathbf{\hat{Y}}_\mathrm{self}\right)
\end{equation}
\begin{equation}
 \mathcal{L}_{\mathrm{cross}}= \frac{1}{N} \sum_{i=1}^{{N}} \xi \left(\mathbf{X}_{\mathrm{cross}},\mathbf{\hat{Y}}_\mathrm{cross}\right)
\end{equation}
\begin{equation}
 \mathcal{L}_{\mathrm{sem}}= \alpha  \mathcal{L}_{\mathrm{self}}+ (1-\alpha)\mathcal{L}_{\mathrm{cross}}\label{eq7}
\end{equation}
where $N$ represents the number of extracted feature blocks. $\xi$ is the cosine similarity distance. 
$\alpha$ is a hyperparameter that controls the intensity of self-domain matching and cross-domain matching.


We follow adversarial training with the Generator and Discriminator to train our model. We express the objective as:
\begin{equation}
    \begin{aligned}
\mathcal{L}_{D}=&-{E}_{\mathbf{y} \sim p_{\text {data}}(\mathbf{y})}\left[\log \mathcal{D}(\mathbf{y})\right] \\
&-{E}_{\mathbf{x} \sim p_{\text {data}}(\mathbf{x})}\left[\log \left(1-\mathcal{D}(\mathcal{G}(\mathbf{x}))\right)\right]
\cr
\end{aligned}
\end{equation}

\begin{equation}
    \begin{aligned}
\mathcal{L}_{\mathcal{G}}=& {E}_{\mathbf{x} \sim p_{\text {data}}(\mathbf{x})}\left[\log \left(1-\mathcal{D}(\mathcal{G}(\mathbf{x}))\right)\right] + \lambda \cdot \mathcal{L}_{\mathrm{sem}}\label{eq9}
\cr
\end{aligned}
\end{equation}
where $\mathcal{L}_{\mathrm{sem}}$ is the multi-scale domain matching loss, and a hyperparameter $\lambda$ is used to control the semantic loss contribution. Our optimization goal is to increase $\log \mathcal{D}(\mathbf{y})$ for real images and decrease $\mathcal{D}(\mathcal{G}(\mathbf{x}))$ for simulation images, resulting in realistic X-ray image generation.

\section{Experiments} \label{human}

\subsection{Experimental Setup}

\textbf{Setup.}
We create a new dataset with unpaired simulation images and real X-ray images. We use CathSim~\cite{https://doi.org/10.48550/arxiv.2208.01455} to capture $2000$ simulation images, and collect $1607$ real X-ray images using the C-arm (Siemens, Germany) and two vascular soft silicone phantoms (Elastrat, Switzerland). We set the hyperparameter $\alpha$ in Equation~\ref{eq7} to $0.5$ to give equal weight to cross-domain and self-domain factors. The term $\lambda$ in Equation~\ref{eq9} is set to $8$. Please refer to the Supplementary Material for more details about our dataset and implementation.




\begin{table}[t]
\centering
\caption{Performance of different methods. Training time is second/epoch.}
\label{tb_test}
\resizebox{\linewidth}{!}{
\setlength{\tabcolsep}{0.25 em} 
\renewcommand{\arraystretch}{1.1}
\begin{tabular}{l|c|c|c}
\hline 			
\textbf{Method}   & \textbf{FID Score }    &  \textbf{\#Param }      &  \textbf{Training Time} \\
\hline 	

CycleGAN \cite{zhu2017unpaired}				         & 337.80 & 28.27M  & 220.8   \\
GcGAN \cite{fu2019geometry}         &  285.37 & 28.27M     & 131.2   \\
FaseCUT \cite{park2020contrastive}        & 297.09 & 14.70M   & 101.0  \\
FLSeSim \cite{zheng2021spatially} 	                  & 307.09 & 14.68M   & 448.3   \\
\hline
Sim2Xray (w/o self-domain)            & 128.84 & \textbf{12.55M}  & \textbf{95.8}  \\
Sim2Xray (w/o cross-domain)            & 138.68 & \textbf{12.55M}  & 96.4  \\
Sim2Xray (ours)        & \textbf{115.77} & \textbf{12.55M}   & 96.1   \\

\hline 	
\end{tabular}
}
\end{table}

\textbf{Baselines.}
We compare our simulation to X-ray (Sim2Xray) method with several recent works, based on visual effects, FID scores~\cite{heusel2017gans}, training time, and the number of parameters used. The compared methods include CycleGAN~\cite{zhu2017unpaired}, GcGAN~\cite{fu2019geometry}, FaseCUT~\cite{park2020contrastive} and FLSeSim~\cite{zheng2021spatially}. We do not compare our approach with pix2xray~\cite{Haiderbhai2020pix2xrayCR} although both methods do X-ray image translation. This is because we used unpaired data while pix2xray~\cite{Haiderbhai2020pix2xrayCR} used paired data and there is no public source code of pix2xray~\cite{Haiderbhai2020pix2xrayCR} for testing.


\subsection{Results}

The results of our model and other methods are shown in Table~\ref{tb_test}. The results show that our proposed method outperforms other approaches significantly in FID scores. In addition, our method is the smallest model in terms of the number of parameters. 
Furthermore, our training time is much less than other methods and our inference time is only $11$ms for each image, compared to around $16$ms of other methods.

\begin{figure*}[h]
\Large%
   \centering
\resizebox{\linewidth}{!}{
\setlength{\tabcolsep}{2pt}
\begin{tabular}{ccccccc}

\shortstack{\includegraphics[width=0.33\linewidth]{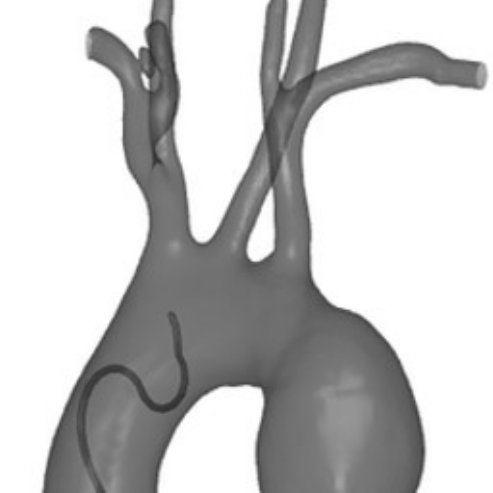}}&
\shortstack{\includegraphics[width=0.33\linewidth]{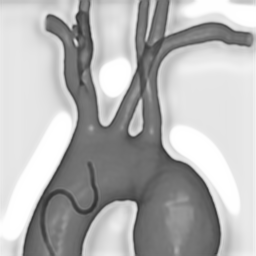}}&
\shortstack{\includegraphics[width=0.33\linewidth]{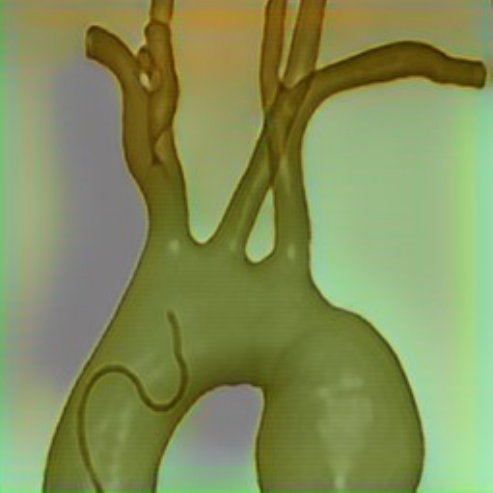}}&
\shortstack{\includegraphics[width=0.33\linewidth]{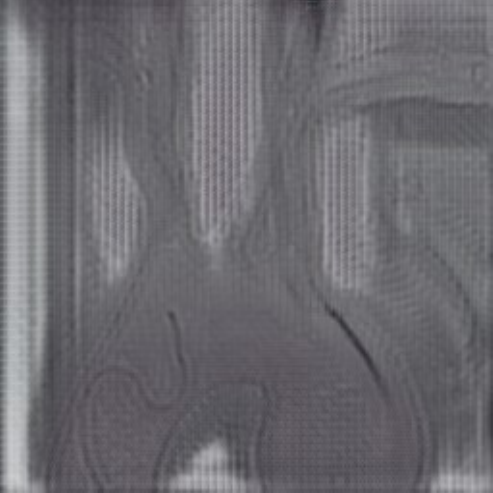}}&
\shortstack{\includegraphics[width=0.33\linewidth]{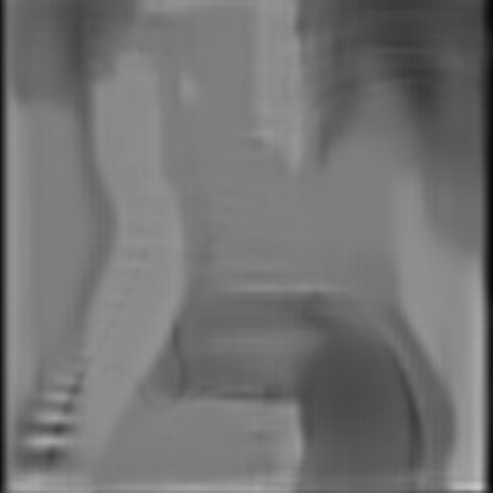}}&
\shortstack{\includegraphics[width=0.33\linewidth]{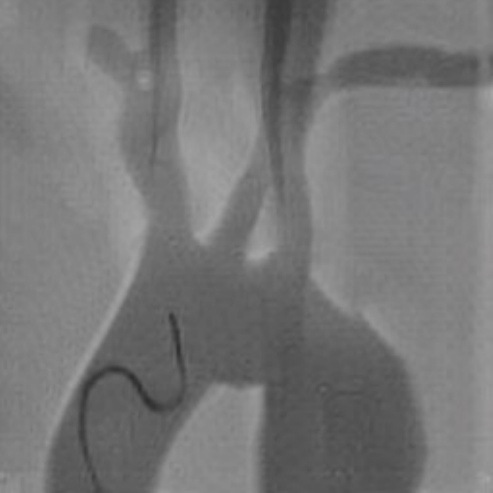}}\\[1pt]

\shortstack{\includegraphics[width=0.33\linewidth]{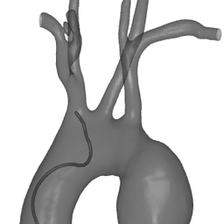}}&
\shortstack{\includegraphics[width=0.33\linewidth]{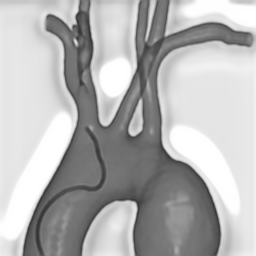}}&
\shortstack{\includegraphics[width=0.33\linewidth]{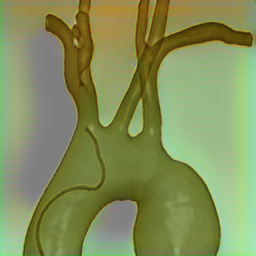}}&
\shortstack{\includegraphics[width=0.33\linewidth]{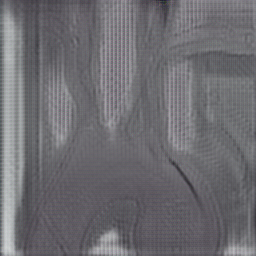}}&
\shortstack{\includegraphics[width=0.33\linewidth]{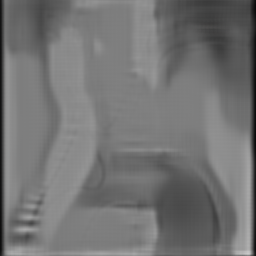}}&
\shortstack{\includegraphics[width=0.33\linewidth]{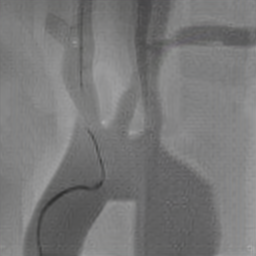}}\\[1pt]

\shortstack{\includegraphics[width=0.33\linewidth]{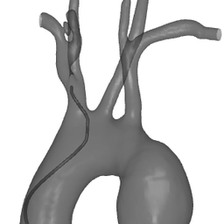}\\ [12pt] (a) Input }&
\shortstack{\includegraphics[width=0.33\linewidth]{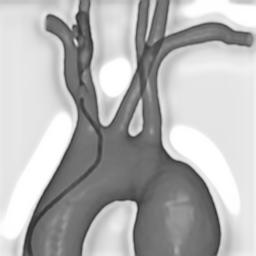}\\ [12pt] (b) FastCUT}& 
\shortstack{\includegraphics[width=0.33\linewidth]{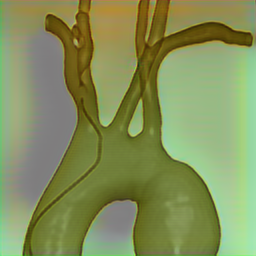}\\ [12pt] (c) FLSeSim}& 
\shortstack{\includegraphics[width=0.33\linewidth]{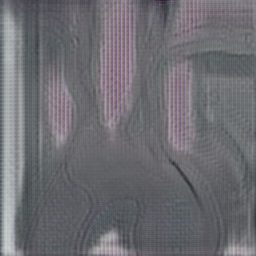}\\ [12pt] (d) GcGAN}& 
\shortstack{\includegraphics[width=0.33\linewidth]{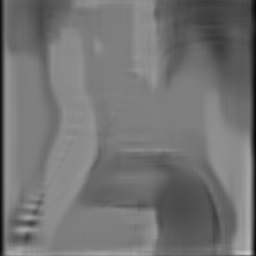}\\ [12pt] (e) CycleGAN}& 
\shortstack{\includegraphics[width=0.33\linewidth]{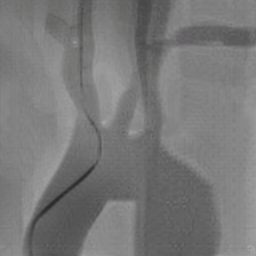}\\ [12pt] (f) Ours}& 
\\[22pt]
     
\end{tabular}
}

 \vspace{-2ex}
 \caption{The translation results of different models.} 
    \label{fig:imgs}
\label{visual}
\end{figure*}

In Fig.~\ref{fig:imgs}, we show qualitative comparisons of our method and all other models. It can be seen that some models such as CycleGAN~\cite{zhu2017unpaired} and GcGAN~\cite{fu2019geometry} have pattern collapse problems. The generated results do not correspond to the input images and do not learn the structural information we expect. FastCUT~\cite{park2020contrastive} and FLSeSim~\cite{zheng2021spatially} do not successfully learn the target domain style of the X-ray images. On the other hand, our method successfully transfers the style of the simulation image into the X-ray image while still retaining the structure information of the input images.  


\begin{figure*}[t]
\Large%
   \centering
\resizebox{\linewidth}{!}{
\setlength{\tabcolsep}{2pt}

\begin{tabular}{ccccccc}


\shortstack{\includegraphics[width=0.4\linewidth]{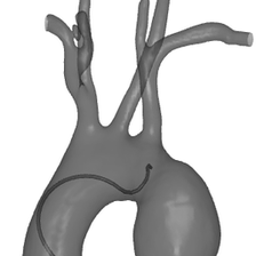}\\  (a) Input}& 
\shortstack{\includegraphics[width=0.4\linewidth]{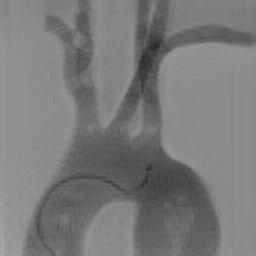}\\  (b) w/o cross-domain}& 
\shortstack{\includegraphics[width=0.4\linewidth]{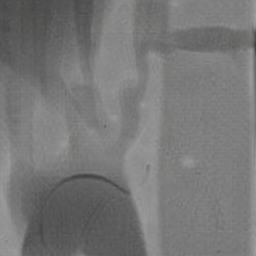}\\  (c) w/o self-domain}& 
\shortstack{\includegraphics[width=0.4\linewidth]{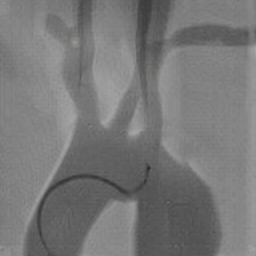}\\  (d) Ours}& 
\\& 
     
\end{tabular}
}
 \vspace{-3ex}
\caption{The contribution of multi-scale semantic matching to the translation results. (a) The input simulation image. (b) The results without using cross-domain matching, (c) The results without using self-domain matching, and (d) Our final results.} 
    \label{fig_intra_inter}
\label{visual}
\end{figure*}

\subsection{Multi-Scale Semantic Matching Analysis}
In Table~\ref{tb_test}, we also demonstrate the contribution of our proposed multi-scale semantic matching. Based on the FID score, we can see that when we only use self-domain or cross-domain matching, the FID score is improved but not optimal. We obtain the best FID score when we use both self-domain and cross-domain matching. It confirms the importance of learning both the style information and keeping the structure information from the input. In Fig.~\ref{fig_intra_inter}, we visualize the results of our model when we do not use self-domain matching or cross-domain matching. From the visualization, we see that the results without cross-domain matching have poor X-ray style, while the results without self-domain matching cannot keep the structure information from the input image. 


\section{Conclusions}\label{Sec:con}
We propose a new and effective method to translate simulation images of an endovascular simulator to X-ray images using multi-scale semantic matching. Our approach has fast training and inference time, making it well-suitable for real-time endovascular simulators. Additionally, we introduce a new dataset that can be used to develop and evaluate image translation models. Our source code and dataset will be made publicly available for future study.



\bibliographystyle{splncs04}
\bibliography{sn-bibliography}

\end{document}